\documentclass[12pt]{iopart}   
\usepackage{graphicx}
\usepackage{epsfig}
\usepackage{amssymb}

\newcommand{\be}{\begin{equation}}
\newcommand{\ee}{\end{equation}}
\newcommand{\beqn}{\begin{eqnarray}}
\newcommand{\eeqn}{\end{eqnarray}}
\begin{document}

\title[Entanglement across extended random defects in the XX spin chain]{Entanglement across extended random defects in the XX spin chain}
\author{R\'obert Juh\'asz}
\address{Wigner Research Centre for Physics, Institute for Solid State
Physics and Optics, H-1525 Budapest, P.O.Box 49, Hungary}
\ead{juhasz.robert@wigner.mta.hu} 
\date{\today}

\begin{abstract}
We study the half-chain entanglement entropy in the ground state of the spin-1/2 XX chain across an extended random defect, where the strength of disorder decays with the distance from the interface algebraically as $\Delta_l\sim l^{-\kappa}$.
In the whole regime $\kappa\ge 0$, the average entanglement entropy is found to increase 
logarithmically with the system size $L$ as $S_L\simeq\frac{c_{\rm eff}(\kappa)}{6}\ln L+const$, where the effective central charge $c_{\rm eff}(\kappa)$ depends on $\kappa$. 
In the regime $\kappa<1/2$, where the extended defect is a relevant perturbation, the strong-disorder renormalization group method gives
$c_{\rm eff}(\kappa)=(1-2\kappa)\ln2$, while, in the regime $\kappa\ge 1/2$, where the extended defect is irrelevant in the bulk, numerical results indicate a non-zero effective central charge, which increases with $\kappa$. The variation of $c_{\rm eff}(\kappa)$ is thus found to be non-monotonic and discontinuous at $\kappa=1/2$.
\end{abstract}

\maketitle

\section{Introduction}
\label{sec:intro}

The entanglement in the ground state of extended quantum systems is an intensively studied problem \cite{amico,entanglement_review,laflorencie}. Among these systems, perhaps the most clearly understood are the entanglement properties of one-dimensional lattice models. A frequently studied measure of entanglement is the entanglement entropy of a subsystem consisting of $L$ consecutive sites, which is the von Neumann entropy of the reduced density matrix $\rho_L$ of the subsystem:
\be  
S_L=-{\rm tr}(\rho_L\ln\rho_L). 
\label{ee}
\ee
For non-critical chains, the entanglement entropy remains bounded with increasing block size $L$, which is a special case of ``area law'' \cite{area}, but in critical points, it increases logarithmically as 
\be 
S_L=a\frac{c}{6}\ln L+const.
\ee
For conformally invariant models, $c$ is the central charge of the corresponding conformal algebra, and as such it is universal \cite{holzhey,vidal,Calabrese_Cardy04}, and $a$ is the number of boundary points of the subsystem.  
In many spatially inhomogeneous variants of the above class of models, this logarithmic dependence still holds in the critical point for the average entanglement entropy, however, $c$ is replaced by an ``effective central charge'' $c_{\rm eff}$, which is characteristic for the model and the type of inhomogeneous modulation. This is the case for the antiferromagnetic spin-$\frac{1}{2}$ XXZ chain with random \cite{refael,Laflo05,fagotti,huse} or aperiodic \cite{aperiodic} couplings, where the ground state is a random \cite{fisherxx} or aperiodic \cite{hida} singlet state, respectively.  
Besides extended inhomogeneities which are present overall in the system, even local defects are able to alter the asymptotics of the entanglement entropy, provided these defects are localised at the boundaries of the subsystem. For the XXZ chain, defect couplings at the boundaries result in an effective cut ($c_{\rm eff}=0$) in the antiferromagnetic regime of the model, and are irrelevant ($c_{\rm eff}=c=1$) in the ferromagnetic regime \cite{zpw}. For the special case of the XX chain, such a defect is a marginal perturbation resulting in a continuous dependence of the effective central charge on the defect coupling \cite{defect}.    
As opposed to a homogeneous background, a local defect (weak coupling) on the top of independent, identically distributed (IID) randomness in the antiferromagnetic XXZ chain is irrelevant in the sense that $c_{\rm eff}=\ln 2$ remains unaltered \cite{vasseur}. 
The effect of an extended, non-local defect \cite{hvl,bariev} on the entanglement entropy scaling has been studied in Ref. \cite{iszl}. Here, a critical transverse-field Ising chain was considered with a perturbation in the coupling strength that decays as $\sim A/l$ with the distance $l$ from the subsystem boundaries and makes the system slightly off-critical. Regarding the local magnetisation at the center of the defect, this kind of perturbation is marginal \cite{bariev} and results in critical exponents varying with $A$ for $A<0$ and local ordering for $A>0$ \cite{ibt,ipt}. The bipartite entanglement entropy is found, however, to saturate to finite values for $A\neq 0$, which means formally that $c_{\rm eff}=0$ \cite{iszl}.  
The entanglement entropy of the transverse-field Ising chain is closely related to that of an XX chain for general parameters \cite{peschel_schotte,IJ07}. 
According to this, the above perturbation is translated in the XX chain as a dimerization of decaying strength on either sublattice depending on the sign of $A$.

In this work, we study the ground state entanglement across an extended random defect in the critical XX chain. This means that the couplings are quenched random variables, but are non-identically distributed, in such a way that the strength of disorder decays algebraically with the distance from the boundary points as $\Delta_l\sim l^{-\kappa}$ \footnote{The surface critical behaviour of the transverse-field Ising chain with a similar type of extended surface disorder has been studied in Ref. \cite{tki}.}. 
This kind of extended defect differs in two respects from the one studied in the transverse-field Ising chain in Ref. \cite{iszl}. First, the perturbation is random, second, the system is not detuned from criticality (on average) even close to the center of the defect. The limiting case $\kappa=0$ corresponds to the random XX chain with identically distributed disorder, where the ground state is a random singlet phase \cite{fisherxx} and the effective central charge is 
$c_{\rm eff}=\ln 2$ \cite{refael}. 
In the other limiting case $\kappa\to\infty$, we obtain an XX chain with local random defects at the boundaries of the subsystem. The effective central charge of the single defect problem is given in terms of an integral which cannot be evaluated for general defect couplings \cite{defect}, nevertheless, it is not greater than $c=1$ for any defect coupling, and so does its average over random couplings.
For intermediate values of $\kappa$, $0<\kappa<\infty$, this kind of defect smoothly interpolates between a disordered environment near the boundaries of the subsystem and a homogeneous bulk asymptotically far away from them, which regions are characterised by different (effective) central charges. 
The entanglement entropy across a similar type of extended defect has been studied recently in a composite system consisting of a random IID subsystem ($\kappa=0$) and extended interface defects ($\kappa>0$) on the other side of boundary points \cite{jkri}. 
Our main interest is the scaling of the average entanglement entropy across the defect for intermediate values of the decay exponent $\kappa$.    
We will also investigate the end-to-end correlations in open chains, which, together with the end-to-end concurrence has been studied in non-random variants of the XX chain from the perspective of exploiting spins chains as channels for quantum teleportation \cite{venuti}.
We will apply a strong-disorder renormalization group (SDRG) method \cite{mdh,fisherxx,im} to the model and study it numerically by means of free-fermion techniques \cite{lschm,peschel03,vidal,jin_korepin}. 

The rest of the paper is organised as follows. In section \ref{sec:model}, the model is defined. In section \ref{sec:model}, the end-to-end correlations and the entanglement entropy are studied in the frame of the SDRG approach, while numerical results obtained by the free-fermion mapping are presented in section \ref{sec:num}. Finally, the results are discussed in section \ref{sec:discussion}.

\section{The XX chain with an extended random defect}
\label{sec:model}

\begin{figure}[h]
\begin{center}
\includegraphics[width=10cm]{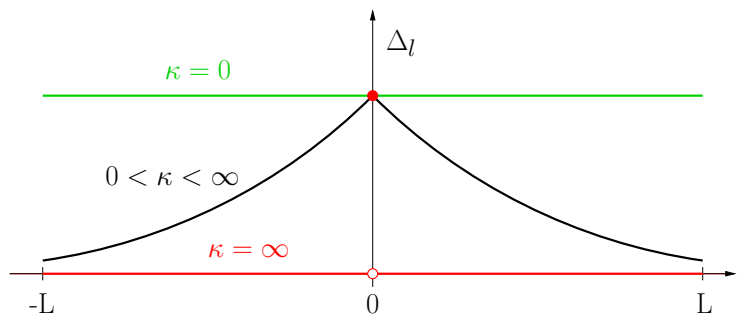} 
\caption{\label{delta} Illustration of the variation of the strength of disorder $\Delta_l$ with the position $l$ in the chain for different values of $\kappa$.
The half-chain entanglement entropy between the subsystems with sites $l>0$ and  $l\le 0$ is considered.  
}
\end{center}
\end{figure}

We will consider the antiferromagnetic XX chain having the Hamiltonian 
\be 
H=\sum_{l=-L+1}^{L-1}J_l(S_l^xS_{l+1}^x+S_l^yS_{l+1}^y),
\label{hamilton}
\ee
where $S_{l}^x$ and $S_{l}^y$ are spin-$\frac{1}{2}$ operators at site $l$. The random couplings $J_l$ are drawn from site-dependent distributions, which are uniform for each $l$ but their support 
$[\frac{1}{2}-\Delta_l,\frac{1}{2}+\Delta_l]$ is shrinking with the distance from the middle of the chain as 
\be 
\Delta_l=\frac{1}{2}(|l|+1)^{-\kappa},
\label{dist}
\ee   
see Fig. \ref{delta}.
This means that the central coupling $J_0$ is uniformly distributed in the interval $[0,1]$, while $J_{L-1}$ is concentrated around $1/2$ in the 
limit $L\to\infty$ (provided that $\kappa>0$). 
We are interested in the entanglement entropy of the subsystem consisting of spins with $l>0$ in the ground state $|\psi\rangle$ of the chain. This is defined through the reduced density matrix $\rho_{L}={\rm tr}_{l\le 0}|\psi\rangle\langle\psi|$ of the subsystem as given in Eq. (\ref{ee}).
Besides open chains, we will also consider periodic ones, which comprise two open segments with different random realisations of couplings as given in Eq. (\ref{hamilton}) and whose end spins are connected with couplings $J_{-L}$ and $J_{L}$ following the distribution specified by Eq. (\ref{dist}). In this case, the subsystem is of size $2L$ and consists of the spins with indices $l>0$ of both segments.  

The Hamiltonian can be mapped to a system of free fermions by standard methods \cite{lschm}. 
The first step of this is to write the Hamiltonian as a fermion chain 
\be 
H=\frac{1}{2}\sum_{l=-L+1}^{L-1}J_l(c_l^{\dagger}c_{l+1}+c_{l+1}^{\dagger}c_{l}),
\ee
by the help of a Jordan-Wigner transformation 
\be 
S_l^x+iS_l^y=c_l^{\dagger}e^{i\pi\sum_{j<l}c_j^{\dagger}c_j}, 
\quad 
S_l^x-iS_l^y=e^{-i\pi\sum_{j<l}c_j^{\dagger}c_j}c_l,
\ee
which can be subsequently diagonalised. 
The reduced density matrix of the subsystem can be written as the exponential of a free-fermion operator \cite{peschel03} and the entanglement entropy can be computed from the eigenvalues of the correlation matrix 
\be 
C_{ij}=\langle c_i^{\dagger}c_j\rangle
\ee
restricted to the subsystem \cite{vidal,peschel03}. 
We are also interested in the correlation of end spins
\be 
G_L^{\alpha}\equiv \langle S_{-L+1}^{\alpha}S_{L}^{\alpha}\rangle,
\ee
$\alpha=x,y,z$, which can be expressed by the correlation matrix element
$C_L\equiv C_{-L+1,L}$ as \cite{venuti}
\beqn
G_L^{x}=G_L^{y}&=&-\frac{1}{2}e^{i\pi L/2}C_{L}, \nonumber \\
G_L^{z}&=&-(C_{L})^2.
\eeqn

\section{Strong-disorder renormalization group}
\label{sec:sdrg}

\begin{figure}[h]
\begin{center}
\includegraphics[width=5.5cm]{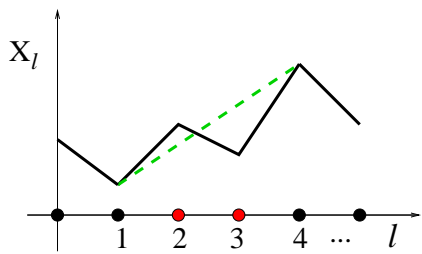} 
\caption{\label{zigzag} The transformation of the zig-zag path under a renormalization step. The strongest coupling is the one connecting site $2$ and $3$. The section of the path between site $1$ and $4$ is replaced by a straight segment.    
}
\end{center}
\end{figure}

The SDRG method is a real-space renormalization scheme which provides an approximate ground state and the low-energy properties of the random XX chain \cite{mdh,fisherxx,im}. In this procedure, the largest coupling $J_m$ in the chain is picked, and the state of the block comprising spins $m$ and $m+1$ is projected onto a singlet, which is a good approximation if $J_m\gg J_{m-1},J_{m+1}$. 
The neighbouring spins $m-1$ and $m+2$ are connected by a weak, effective antiferromagnetic coupling $\tilde J=J_{m-1}J_{m+1}/J_m$ obtained by second-order perturbation theory. Applying this step iteratively, the largest coupling, 
which sets the renormalization scale $\Omega=\max\{J_l\}$, is gradually decreased and the approximation becomes more and more accurate so that, in an infinite system, it becomes asymptotically exact in the so called infinite-randomness fixed point (IRFP) of the transformation. The resulting ground state is a product of singlet states of pairs of spins, which is called random singlet state \cite{fisherxx}.   
The procedure can be graphically represented by introducing a zig-zag path 
\beqn 
&X_{-L}&=0  \nonumber \\   
&X_l&=\sum_{n=-L+1}^l(-1)^n\ln(\Omega_0/J_n), \quad -L<l<L \nonumber  \\
\label{Xdef}
\eeqn
which is associated with a given finite realisation of the open chain. Here, $\Omega_0=\sup\{J_0\}=1$. 
The elimination of the largest coupling corresponds to the smoothing out of the smallest valley as illustrated in Fig. \ref{zigzag}. 
This formulation of the SDRG procedure is formally identical to the renormalization of the potential landscape of the Sinai walk \cite{sinai}. 

This representation can be used to infer the finite-size scaling of the energy gap $\epsilon_L=\tilde J_L/2$, where $\tilde J_L$ is the strength of the last effective coupling generated by the SDRG procedure. Due to the simple rule illustrated in Fig. \ref{zigzag}, the strength of an effective coupling between spin $i$ and $j$ is related to the shift of the path as 
$|X_i-X_j|=\ln(\Omega_0/\tilde J_{ij})$. 
It is easy to see that $\{X_{2l}\}_{l=-L/2}^{L/2-1}$ is a random walk in discrete time $l=-L/2,-L/2+1,\dots,L/2-1$, and $|X_i-X_j|$ is just the displacement of the walk after $|i-j|$ steps.  
As the length of the lastly generated effective coupling in a finite system is $\mathcal{O}(L)$, we obtain in the case of IID couplings ($\kappa=0$)
the relation $|X_i-X_j|\sim \sqrt{L}$, and the well-known finite-size scaling of the energy gap
$|\ln\epsilon_L|\sim \sqrt{L}$.
This result can be easily generalised for $\kappa>0$. 
In that case, starting at $l=0$ and following the path $X_l$ in either direction, we have a random walk where the step length is shrinking in time as $\sigma_l\sim |l|^{-\kappa}$. The variance of the displacement after $\mathcal{O}(L)$ steps is $\sigma^2(L)=\sum_{l=0}^{\mathcal{O}(L)}\sigma_l^2$. 
For $\kappa<1/2$, this yields $\sigma^2(L)\sim L^{1-2\kappa}$, and we obtain for the finite-size scaling of the energy gap
\be 
|\ln\epsilon_L|\sim L^{\frac{1}{2}-\kappa}  \qquad (\kappa<\frac{1}{2}).
\label{gap}
\ee
For the case $\kappa=1/2$, we obtain $\sigma^2(L)\sim \ln L$, which would imply 
$|\ln\epsilon_L|\sim \sqrt{\ln L}$. The gap obtained by the SDRG method in this case is larger than that of the homogeneous system, $\epsilon_L\sim L^{-1}$ \cite{lschm}, which suggests that the randomness is irrelevant for $\kappa=1/2$ (and in the whole range $\kappa\ge 1/2$), meaning that the ground state of the system and local quantities far away from the center of the defect tend to those of the homogeneous system. 
This is in agreement with the conclusions of the analysis of the surface order parameter for a random, extended surface defect in Ref. \cite{jkri}. 
The validity of the SDRG method is thus restricted to the regime $\kappa<1/2$.

\subsection{End-to-end correlations}

Concerning the magnitude of end-to-end correlations of open chains, we must distinguish between two classes of random samples. There are rare samples in which the two end spins form a singlet with each other. In this case, the correlations $|G_L^{\alpha}|$ are $\mathcal{O}(1)$ irrespective of the length of the chain \cite{hoyos}. 
One can easily show that this happens in precisely those samples for which the two end values of the corresponding path $X_{-L}$ and $X_{L-1}$ are the global extremal values of the path, i.e. either $X_{-L}$ is the maximum and $X_{L-1}$ is the minimum (for $L$ even) or vice versa (for $L$ odd). 
In other words, the path has a surviving character at both ends, i.e. it does not cross its initial level $X_{-L}$ and $X_{L-1}$. 
The survival probability $P_s(t)$ of a random walk with identically distributed jump lengths ($\kappa=0$) is known to decrease with the number of steps as $P_s(t)\sim t^{-1/2}$ \cite{redner}, therefore, for large $L$, the fraction of such rare samples is $[P_s(L)]^2\sim\mathcal{O}(L^{-1})$. 

For $\kappa>0$, the asymptotics of the survival probability can be calculated by turning to a continuous-time limit. Then we can write up a diffusion equation for the distribution $p(x,t)$ of the walker,
\be 
\frac{\partial p}{\partial t}=D(t)\frac{\partial^2p}{\partial x^2}
\ee
with a time-dependent diffusion constant
\be 
D(t)=D_0\left[\frac{|L-t|+1}{L}\right]^{-2\kappa},
\ee
in the time interval $1\le t\le 2L-1$. Such a problem can be reformulated as a time-independent diffusion equation in terms of a reduced time which reads in the range $t\le L$ as 
\be 
\tau(t)=\frac{1}{1-2\kappa}[L-(L-t+1)^{1-2\kappa}L^{2\kappa}].
\ee
The reduced time elapsed until $t=2L-1$ is then
\be 
\tau(t=2L-1)=\frac{2}{1-2\kappa}(L-L^{2\kappa}).
\ee
In the range of validity of the SDRG method, $\kappa<1/2$, $\tau(t=2L-1)\sim L$ in leading order, consequently the survival probability decays asymptotically as 
$P_s(L)\sim [\tau(t=2L-1)]^{-1/2}\sim L^{-1/2}$, just as for $\kappa=0$. 
The fraction of rare samples in the whole range $\kappa<1/2$ is therefore 
$\mathcal{O}(L^{-1})$.

The typical situation, which occurs in almost all samples in the limit $L\to\infty$, is, however, that the two end spins do not form a singlet with each other.
 In this case, the magnitude of the correlations is in the order of the effective couplings of length $\mathcal{O}(L)$, i.e. 
\be 
|\ln(|G_L^{\alpha}|)|\sim L^{\frac{1}{2}-\kappa}.   \qquad  (\kappa<\frac{1}{2})
\label{Gscale<}
\ee
Since the typical correlations are very weak, the average correlations are dominated by the contribution of rare samples, which scales as their proportion:
\be 
|\overline{G_L^{\alpha}}|\sim L^{-1}.  \qquad  (\kappa<\frac{1}{2})
\label{G_av}
\ee
For the special case of identically distributed couplings ($\kappa=0$), this result has been known from the analysis of the surface order parameter \cite{ijr}.  

For general values of $\kappa$, the scaling of the correlation $G_L^{\alpha}$ for large $L$ can be inferred via $G_L^{\alpha}\sim \langle S_{-L+1}^x\rangle\langle S_L^x\rangle$, 
where $\langle S_{-L+1}^x\rangle$ and $\langle S_L^x\rangle$ are surface order parameters with fixed spin boundary conditions at the other end of the chain, $S_L^x=\pm\frac{1}{2}$ and $S_{-L+1}^x=\pm\frac{1}{2}$, respectively, which can be calculated exactly \cite{peschel_surf,ijr}. The order parameter $\langle S_{-L+1}^x\rangle$ can be written in terms of the variable $X_l$ defined in Eq. (\ref{Xdef}) as 
\be 
\langle S_{-L+1}^x\rangle=\frac{1}{2}\left[1+\sum_{l=-L/2+1}^{L/2-1}e^{2X_{2l}}\right]^{-1/2},
\label{surf}
\ee
and we have a similar expression for $\langle S_L^x\rangle$.   
In the regime $\kappa<1/2$, the farthest position reached by the random walk in the positive direction is $X^{\rm max}\equiv\max\{X_{2l}\}\sim L^{1/2-\kappa}$ in typical samples, and the corresponding term $e^{2X^{\rm max}}$ in the sum in Eq. (\ref{surf}) dominates over other terms, leading finally to the scaling in Eq. (\ref{Gscale<}). 
In the case $\kappa=1/2$, one can show that the maximal term $e^{2X^{\rm max}}\sim e^{const\cdot \sqrt{\ln L}}$ no longer dominates the sum, and the latter is in leading order proportional to $L$. This leads to the following scaling of typical correlations, which holds also in the regime $\kappa>1/2$: 
\beqn 
|G_L^{x}|=|C_L|\sim L^{-1} \quad (\kappa\ge\frac{1}{2}).
\label{Gscale>}
\eeqn

\subsection{Entanglement entropy}

In the random singlet state, the entanglement entropy of a bipartition is proportional to the number of singlets connecting the two parts of the system, each giving a contribution $\ln 2$. In the case of an infinite system with identically distributed couplings, the average number of singlets crossing a given interface, which are generated during the SDRG procedure up to the logarithmic energy scale $\Gamma=\ln(\Omega_0/\Omega)$ was shown to increase asymptotically as 
\be 
N(\Gamma)=\frac{1}{6}\ln\Gamma + const.
\label{NGamma}
\ee 
in Ref. \cite{refael}. Considering a finite but large subsystem of size $L$, the formation of singlets involving the spins in the subsystem terminates at the scale $\Gamma_L\sim \sqrt{L}$, which, together with Eq. (\ref{NGamma}), gives for the size-dependence of the average entanglement entropy the well-known result $S_L=a\frac{\ln 2}{6}\ln L+const$ \cite{refael}. 
In Ref. \cite{jkri}, it has been shown that a critical system with a spatially varying strength of the disorder renormalizes to an effective system with a homogeneous randomness beyond some finite scale $\Gamma_H$. 
In this effective system, the couplings are identically distributed and their initial position-dependence is transformed to a position-dependence of their length. 
The asymptotic relation in Eq. (\ref{NGamma}), which gives the average number of connecting singlets as a function of $\Gamma$, therefore remains valid for initially position-dependent coupling distributions, as well. 
Renormalizing a finite system with an extended random defect, the logarithmic energy scale goes up to $\Gamma_L\sim L^{\frac{1}{2}-\kappa}$ according to Eq. (\ref{gap}), which, by substituting it in Eq. (\ref{NGamma}), yields for the size-dependence of the average entanglement entropy
\be 
S_L=a(1-2\kappa)\frac{\ln 2}{6}\ln L+const.  \qquad  (\kappa<\frac{1}{2}).  
\ee
The effective central charge
\be 
c_{\rm eff}(\kappa)=(1-2\kappa)\ln 2 \qquad (\kappa<\frac{1}{2}), 
\label{ceffkappa}
\ee
is thus reduced compared to that of the IID random system ($\kappa=0$), and tends to zero as $\kappa\to 1/2$ from below.

\section{Numerical results}
\label{sec:num}

We have calculated the end-to-end correlations and the entanglement entropy numerically by the free-fermion technique in systems of different sizes typically up to $2048$ spins. The number of samples was in most cases $10^6$ but at least a few times $10^5$ for the largest size. 

\subsection{End-to-end correlations}

\begin{figure}[h]
\begin{center}
\includegraphics[width=9cm]{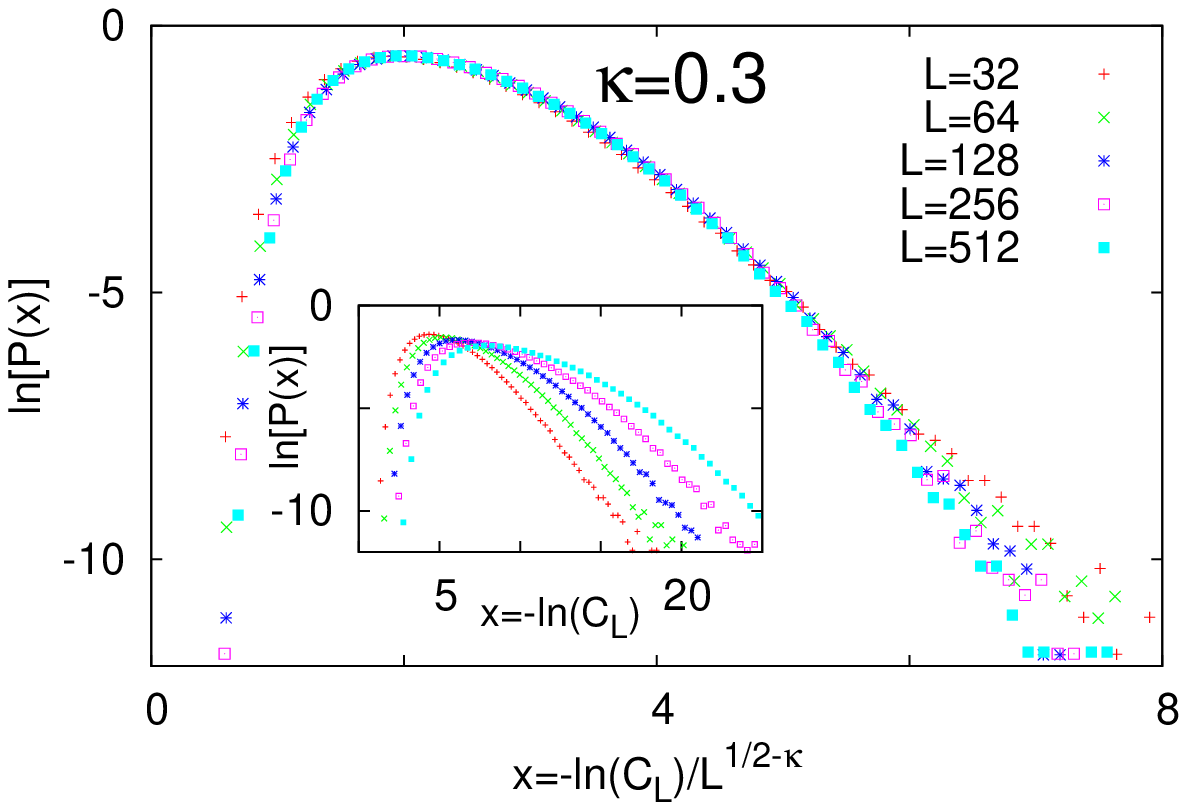}
\includegraphics[width=9cm]{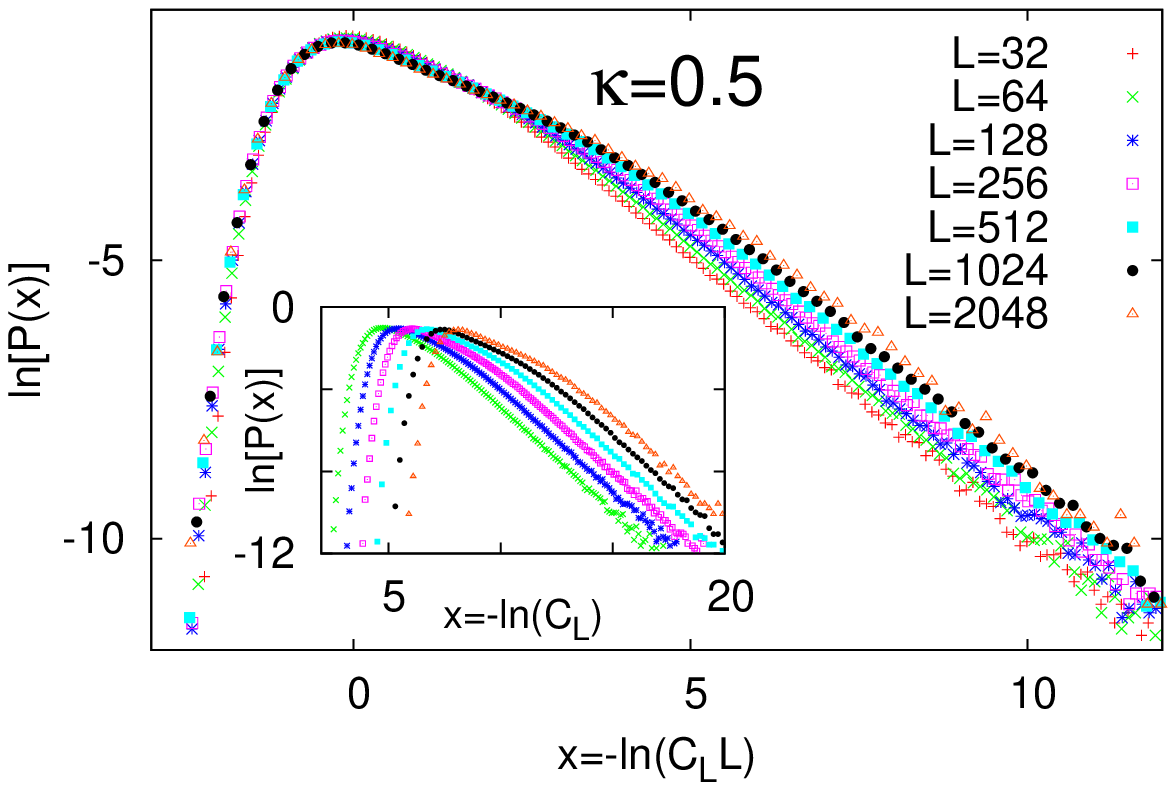}
\includegraphics[width=9cm]{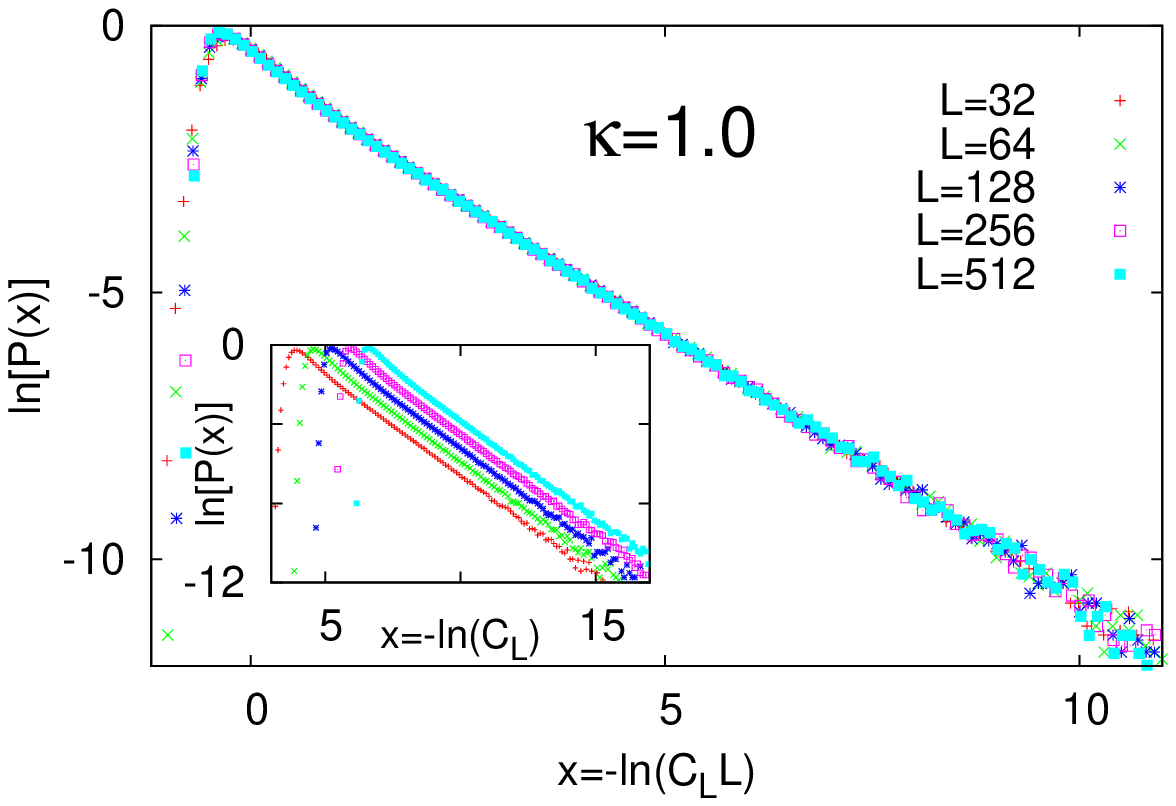}
\caption{\label{cdist}   
The distributions of the end-to-end correlation $C_L\equiv\langle c_{-L+1}^{\dagger}c_L\rangle$ obtained numerically in open systems of different sizes $2L$, for $\kappa=0.3$ (top), $\kappa=0.5$ (middle), and $\kappa=1$ (bottom). The main figures show rescaled distributions, while the insets show the original data. 
}
\end{center}
\end{figure}

\begin{figure}[h]
\begin{center}
\includegraphics[width=8cm]{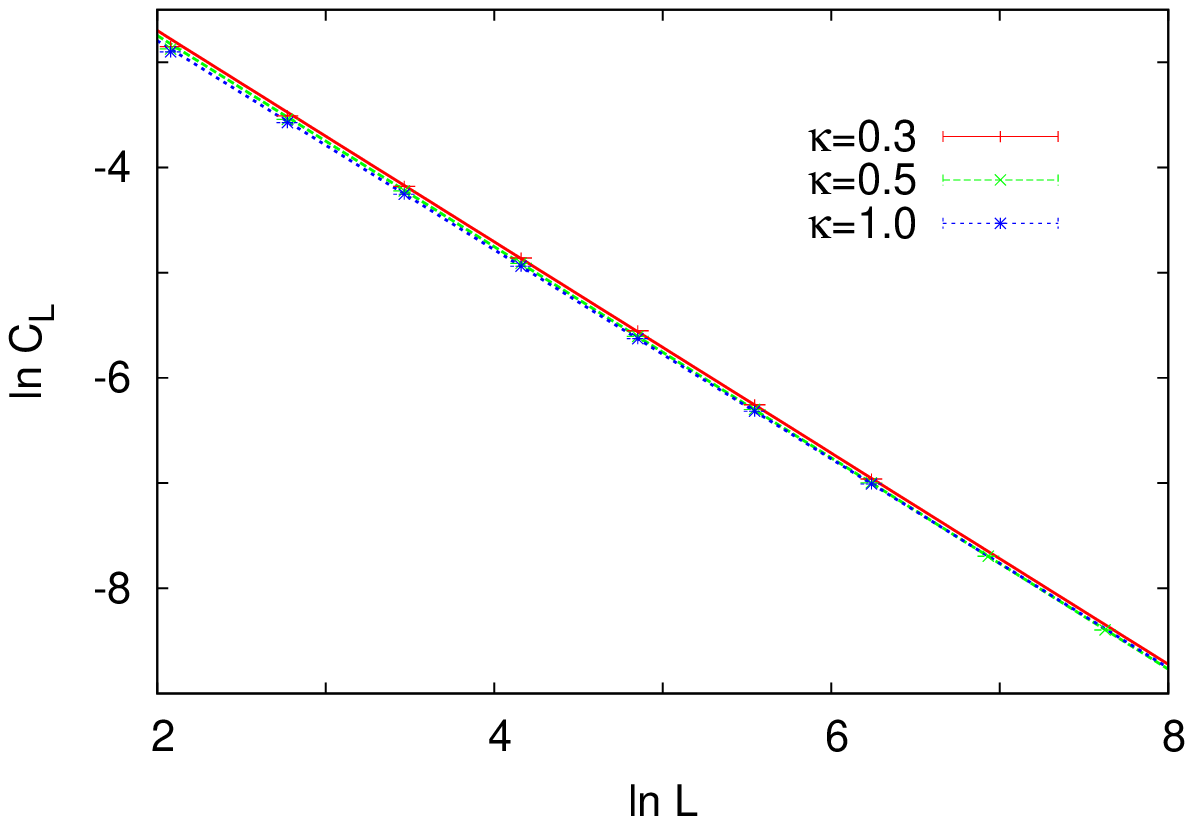}
\caption{\label{c_av}   
Size-dependence of the average end-to-end correlation of fermion operators $\overline{C_L}\equiv\overline{\langle c_{-L+1}^{\dagger}c_L\rangle}$ calculated numerically for different values of $\kappa$. The slope of the straight lines fitted to the data deviate from $-1$ by at most 1\%. 
}
\end{center}
\end{figure}

The distributions of the end-to-end correlation $C_L\equiv\langle c_{-L+1}^{\dagger}c_L\rangle$ in open chains for different sizes are shown Fig. \ref{cdist}. 
In accordance with the prediction of the SDRG method in Eq. (\ref{Gscale<}), the distributions of the scaling variable $\ln(C_L)/L^{1/2-\kappa}$ exhibit a data collapse for $\kappa<1/2$, as it is illustrated in Fig. \ref{cdist} for $\kappa=0.3$.
In the regime $\kappa\ge 1/2$, where the extended defect is irrelevant, the appropriate scaling combination, according to Eq. (\ref{Gscale>}), is $C_LL$. The numerical results shown in Fig. \ref{cdist} for $\kappa=1/2$ and $\kappa=1$ are in agreement with this, although there are strong corrections to scaling at $\kappa=1/2$ in the regime of small correlations.  
The finite-size scaling of the average correlations shown in Fig. \ref{c_av} is found to follow the law $\overline{C_L}\sim L^{-1}$ obtained in the previous section for any $\kappa\ge 0$. 

As we have argued in the previous section, for $\kappa<1/2$, the extended random defect is a relevant perturbation and the critical behaviour of the system is described by an infinite-randomness fixed point. For $\kappa\ge 1/2$, it is, however, an irrelevant perturbation and the critical scaling is the same as that of the clean system. 
We can see that the finite-size behaviour of the end-to-end correlations obtained numerically supports this picture. 

\subsection{Entanglement entropy}

\begin{figure}[h]
\includegraphics[width=8cm]{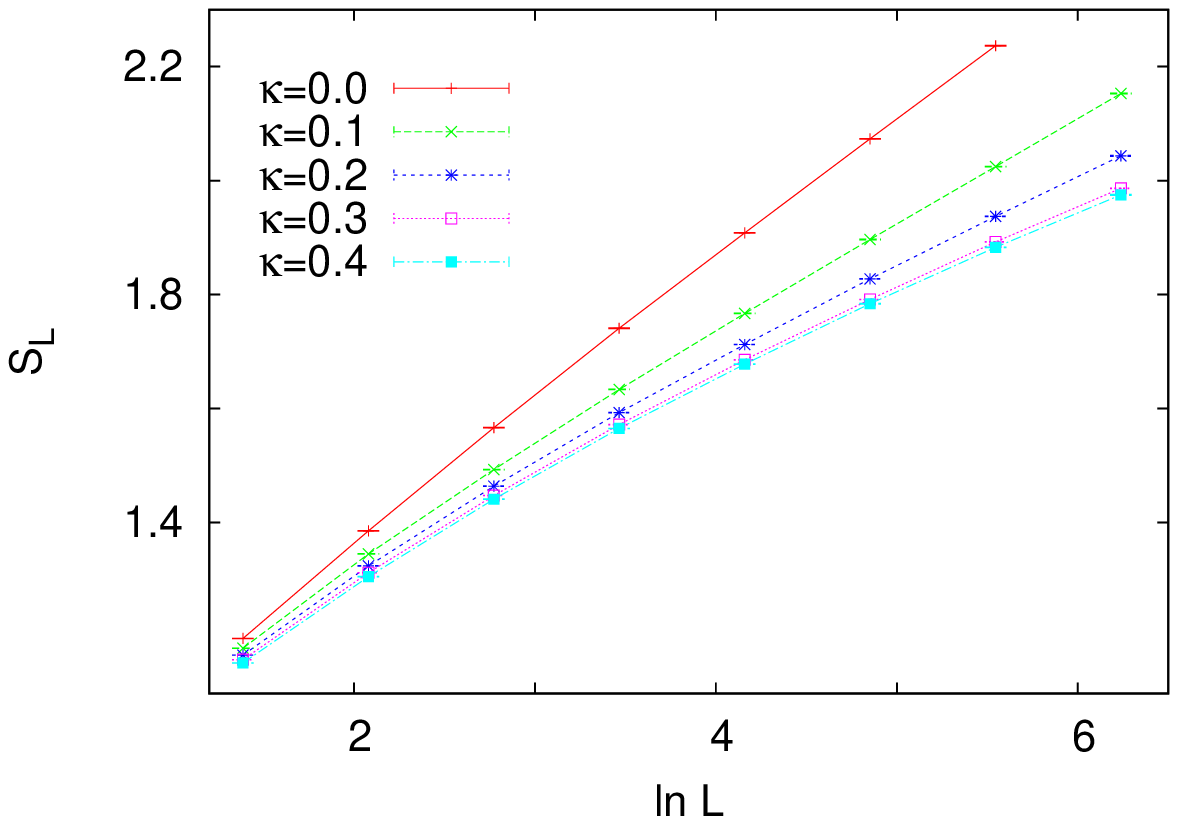}
\includegraphics[width=8cm]{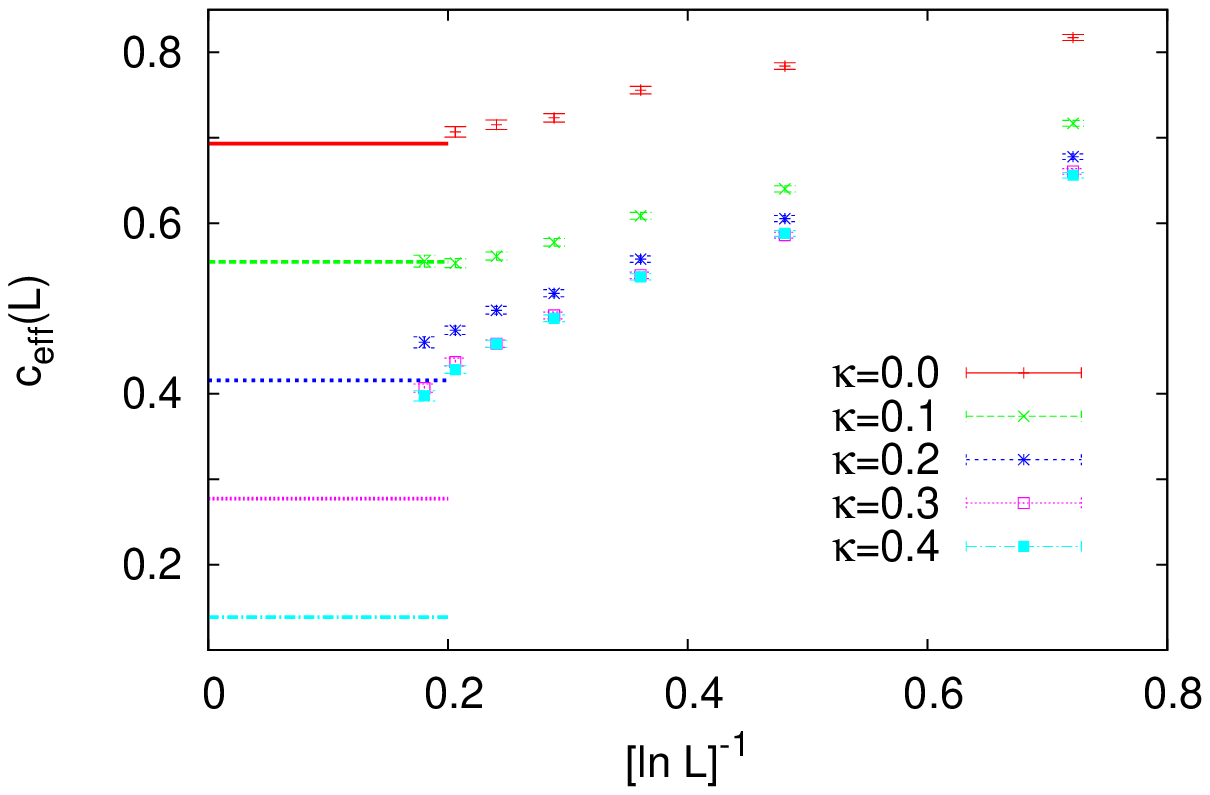}
\caption{\label{kappa<}   
Left. Size-dependence of the average entanglement entropy calculated numerically in the periodic system for different values of $\kappa$. 
Right. The effective, size-dependent central charge defined in Eq. (\ref{ceff}) plotted against $[\ln L]^{-1}$. The horizontal lines indicate the asymptotic values $c_{\rm eff}=(1-2\kappa)\ln 2$ obtained by the SDRG method. 
}
\end{figure}

Next, let us turn to the behaviour of the average entanglement entropy.
The results obtained for different system sizes, together with 
the effective, size-dependent central charges defined as 
\be 
c_{\rm eff}(L)=\frac{6}{a}\frac{S_{2L}-S_L}{\ln 2}.  
\label{ceff}
\ee
are shown in Fig. \ref{kappa<} for different values of $\kappa$ in the range $\kappa<1/2$.
A tendency toward the asymptotic value $c_{\rm eff}=(1-2\kappa)\ln 2$ predicted by the SDRG method can be seen, but, for not too small values of $\kappa$, the finite-size estimates are still far from the asymptotic value. This can be understood, since, according to the SDRG treatment, the extended defect in a system of $L$ spins with some $\kappa<1/2$ renormalizes to an effective IID system of $\mathcal{O}(L^{1-2\kappa})$ spins. Thus the crossover size $L_{\kappa}$, beyond which the asymptotic finite-size dependences characteristic of the IRFP are valid to a good approximation, is $L_{\kappa}\sim L_0^{1/(1-2\kappa)}$. As $\kappa\to 1/2$ this scale diverges and, for $\kappa$ close to $1/2$, it is considerably larger than the system sizes available by our numerical method. 

\begin{figure}[h]
\begin{center}
\includegraphics[width=8cm]{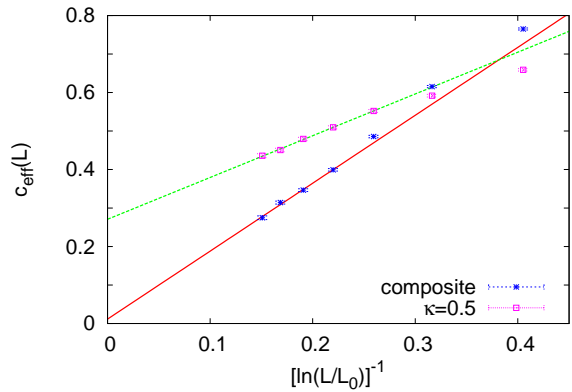}
\caption{\label{kappa=}   
The effective, size-dependent central charge defined in Eq. (\ref{ceff}) plotted against $[\ln(L/L_0)]^{-1}$ with $L_0=0.34$ for $\kappa=1/2$. 
As a comparison, we have replotted the data obtained in the composite system of Ref. \cite{jkri} with $\kappa=1/2$.
}
\end{center}
\end{figure}

\begin{figure}[h]
\includegraphics[width=8cm]{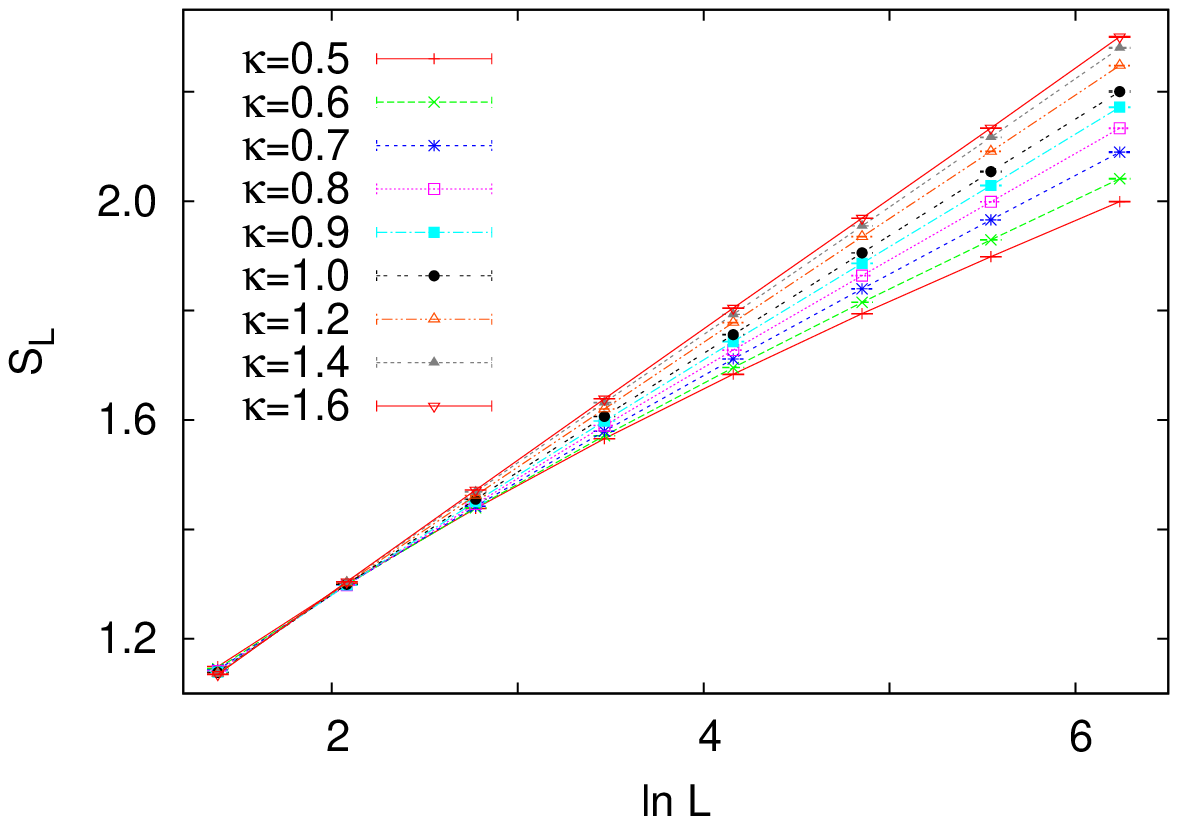}
\includegraphics[width=8cm]{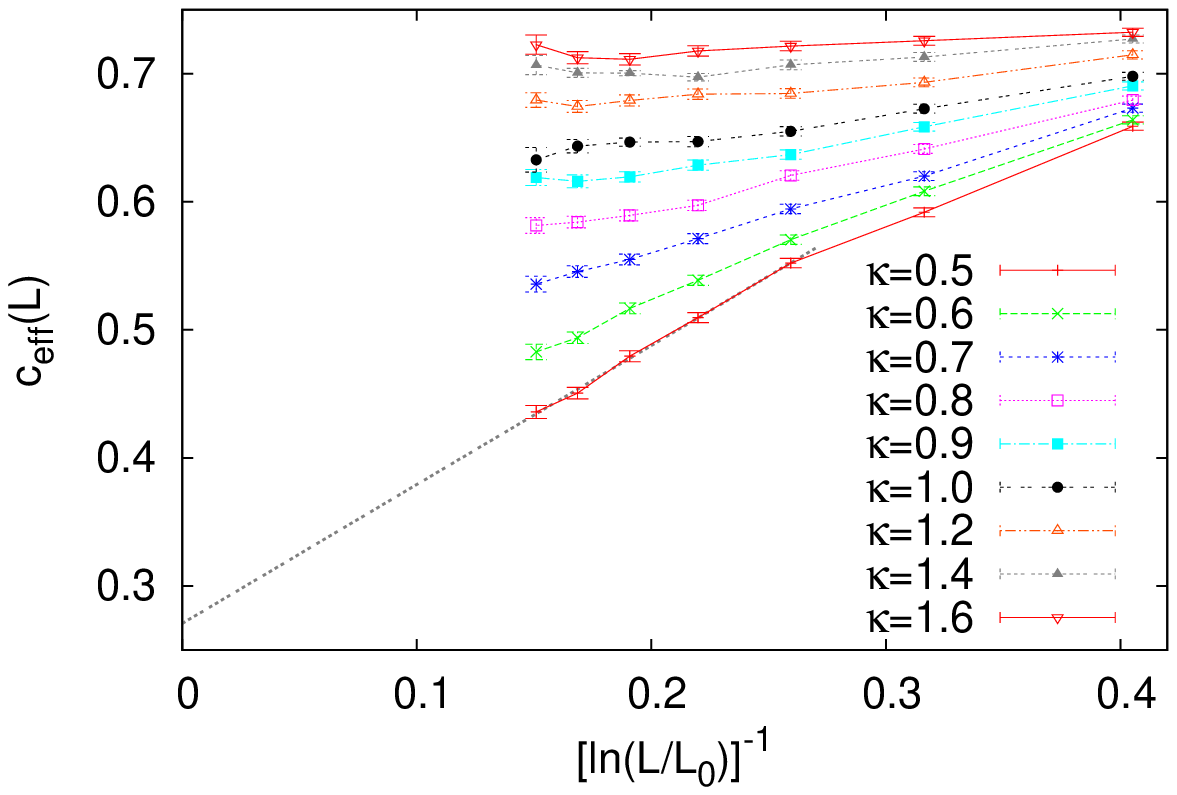}
\caption{\label{kappa>}   
Left. Size-dependence of the average entanglement entropy calculated numerically in the periodic system for different values of $\kappa$. 
Right. The effective, size-dependent central charge defined in Eq. (\ref{ceff}) plotted against $[\ln(L/L_0)]^{-1}$ with $L_0=0.34$. 
}
\end{figure}

\begin{figure}[h]
\begin{center}
\includegraphics[width=8cm]{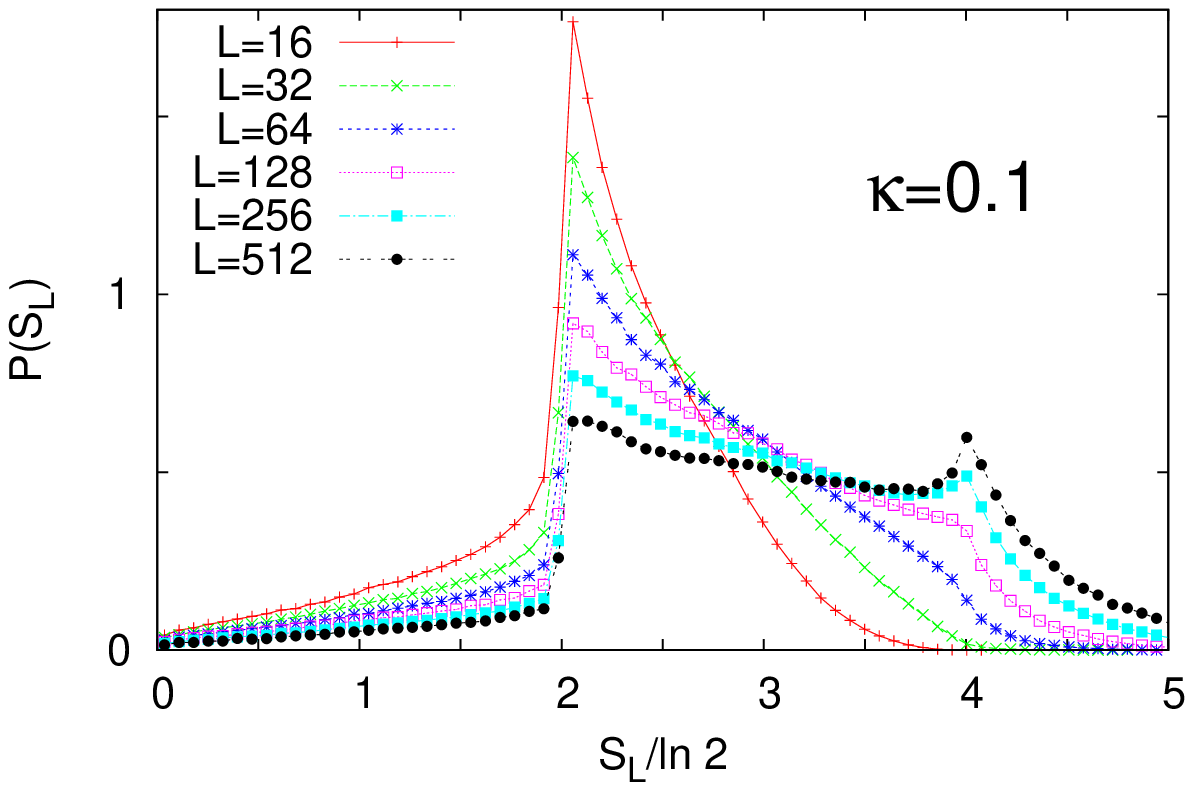}
\includegraphics[width=8cm]{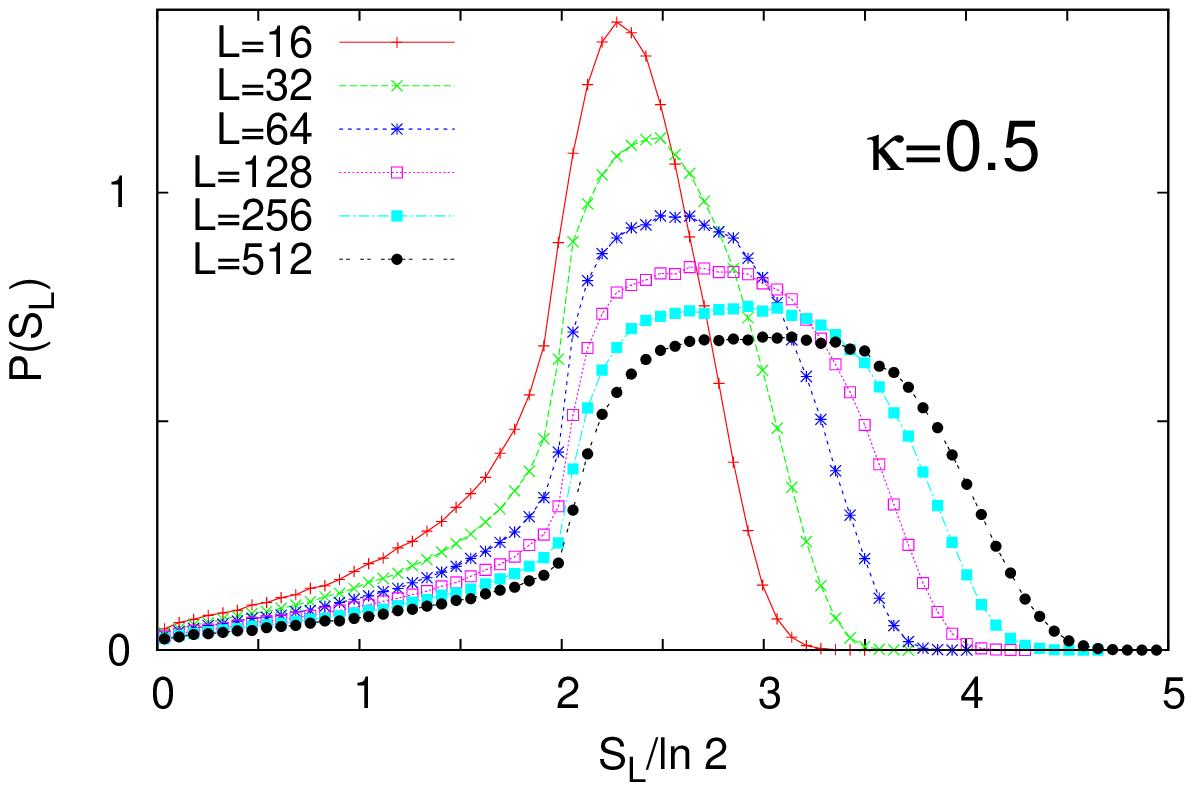}
\includegraphics[width=8cm]{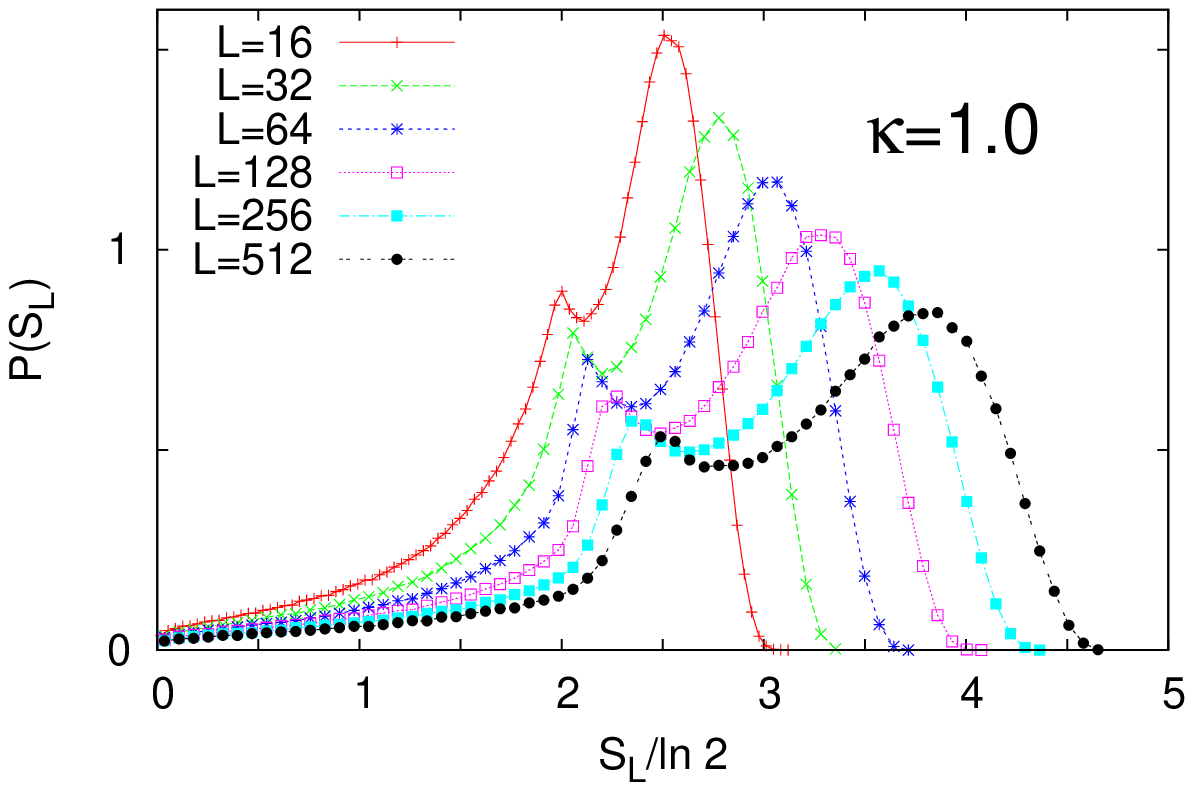}
\caption{\label{sdist}   
Distributions of the sample-dependent entanglement entropies calculated numerically in the periodic system for different system sizes for $\kappa=0.1$ (top), $\kappa=0.5$ (middle), and $\kappa=1$ (bottom). 
}
\end{center}
\end{figure}

The size-dependence of the entanglement entropy and the calculated effective central charges in the range $\kappa\ge 1/2$ can be seen in Fig. \ref{kappa>}. 
The borderline case $\kappa=1/2$ is shown separately in Fig. \ref{kappa=}. 
Here, we have also plotted $c_{\rm eff}(L)$ for the composite system considered in Ref. \cite{jkri}, which is identical to the model studied in this work for $l\ge 0$ but contains IID randomness for $l<0$. In that model, the SDRG approach predicts the same effective central charge as that of the present model for $\kappa\le 1/2$, while, for $\kappa\ge 1/2$,
the average entanglement entropy was found to increase sub-logarithmically as $S_L\sim \ln\ln(L/L_0)$, which is tantamount to a vanishing effective central 
charge as $c_{\rm eff}(L)\sim [\ln(L/L_0)]^{-1}$. 
For the extended defect model with $\kappa=1/2$, the numerical results for the effective central charge are compatible with a linear decrease with $[\ln(L/L_0)]^{-1}$, as well, see Fig. \ref{kappa=}, but the extrapolation to $L\to\infty$ gives a non-zero asymptotic value $c_{\rm eff}(\kappa=1/2)=0.27(4)$. This means a logarithmic increase of the average entanglement entropy at $\kappa=1/2$ with a double-logarithmic subleading term: 
\be
S_L\sim \frac{a}{6}c_{\rm eff}\ln L + c_2\ln\ln(L/L_0).   \qquad (\kappa=\frac{1}{2})
\ee 
For $\kappa\ge 1/2$, the effective central charges for a fixed $L$ are increasing with $\kappa$, see Fig. \ref{kappa>}, and, for large enough $\kappa$, the data seem to saturate to $\kappa$-dependent asymptotic values. This suggests that  
the asymptotic $c_{\rm eff}$ increases with $\kappa$ in this range, and the extended defect acts like a localised defect whose defect strength decreases with increasing $\kappa$.  
The distributions of the sample-dependent entanglement entropies for different sizes are shown in Fig. \ref{sdist}. For $\kappa<1/2$, peaks can be seen to emerge at even multiples of $\ln2$, which indicates that the critical behaviour is controlled by an IFRP and the corresponding ground state (on large scales) is a random singlet state. 
This is no longer the case for $\kappa\ge 1/2$. At the borderline case a roughly flat region appears in the distributions, while, for $\kappa>1/2$, a double-peak shape emerges. 

\section{Discussion}
\label{sec:discussion}

In this work, we have studied the half-chain entanglement entropy and the end-to-end correlations in the XX chain across random extended defects with algebraically decaying strength of disorder. 
The ground state of the system is found to be qualitatively different in the regimes $\kappa<1/2$ and $\kappa\ge 1/2$ of the decay exponent. In the former case, the extended defect is a relevant perturbation, the critical behaviour is controlled by the infinite-randomness fixed point of the SDRG procedure and the ground state is a random singlet state.
This state is, however, different from the ground state of the IID system. Here, due to the position-dependence of the strength of disorder, a local crossover length scale can be defined which increases with the distance $l$ from the center of the defect as $\xi_l\sim l^{2\kappa}$. 
Within this local crossover length scale, the system behaves locally as a clean one, and singlet pairs emerge only beyond $\xi_l$. 
The arrangement of singlets is thus inhomogeneous, as short singlets
up to the length $\xi_l\sim l^{2\kappa}$ are absent in a distance $l$ from the center. This inhomogeneously ``diluted'' structure of singlets can, however, 
be simply transformed to a homogeneous one by changing to a reduced distance $\tilde l\sim l^{1-2\kappa}/(1-2\kappa)$.
In the regime $\kappa\ge 1/2$, the random extended defect is irrelevant far from the center and the local quantities are expected to behave there as in the clean system. 

We have confirmed the correctness of this scenario in both regimes of $\kappa$ by numerically calculating the end-to-end correlation in open chains.  
Concerning the entanglement entropy, which is a non-local quantity, the above considerations about the inhomogeneous random singlet state in the regime $\kappa <1/2$ result in that the effective central charge is reduced by a factor $1-2\kappa$ compared to that of the IID model. 
Thus, although the defect gets less and less extended for increasing $\kappa$, the effective central charge decreases since the singlet structure becomes sparser and sparser.    
In the regime $\kappa\ge 1/2$, the numerical results still indicate a logarithmic increase of the entanglement entropy in leading order but the decreasing tendency of $c_{\rm eff}(\kappa)$ with $\kappa$ is reversed. Here, the ground state far from the center tends to a homogeneous state, so the situation is similar to the case of a local defect \cite{defect}, where any defect coupling irrespective of whether it is smaller or larger than the bulk coupling leads to a reduction of the effective central charge, the extent of which increases with the deviation from the bulk value. In our model, increasing $\kappa$ makes the deviations (apart from the center) smaller and smaller, leading thus to an increasing effective central charge.   
In conclusion, the dependence of $c_{\rm eff}(\kappa)$ on $\kappa$ is non-monotonic and, in addition to this, the numerical results point toward a discontinuity at $\kappa=1/2$, where $\lim_{\kappa-1/2\to 0-}c_{\rm eff}(\kappa)=0\neq c_{\rm eff}(\kappa=1/2)$.

In the definition of the extended defect, we have used a particular form of coupling distributions. The derivation of results in the frame of the SDRG method, in particular the effective central charge in Eq. (\ref{ceffkappa}), however, does not rely on specific properties of the distributions other than the exponent $\kappa$ describing the asymptotic decay of the position-dependent standard deviation of couplings through $\sigma_l\sim |l|^{-\kappa}$. The results in the regime $\kappa<1/2$ are therefore universal in this sense. Keeping in mind the marginal behaviour of the entanglement entropy for a local defect \cite{defect}, this universality is, however, no longer expected to hold in the regime $\kappa\ge 1/2$.

The results obtained here for the XX model in the regime $\kappa<1/2$ are also valid for the antiferromagnetic XXZ chain, for which the SDRG method produces the same (inhomogeneous) random singlet state as for the XX chain \cite{fisherxx}. 
In the regime $\kappa\ge 1/2$, however, the behaviour of the entanglement entropy is expected to be different from that of the XX chain since, here, even a local defect is known to inhibit the unbounded growth of the entanglement entropy \cite{zpw}. 

The model studied in this work and the composite system of Ref. \cite{jkri} can be regarded as particular cases of a more general class of models where the strength of disorder in the two subsystems are allowed to decay with different exponents, $\kappa_+$ and $\kappa_-$. Some of the results obtained here and in Ref. \cite{jkri} can be easily generalised to this class of models. If the extended defect is relevant in both subsystems, i.e. $\kappa_+,\kappa_-<1/2$, the number of connecting singlets is controlled by the subsystem in which the number of active spins at a given (common) scale $\Gamma$ is smaller, so the SDRG method leads ultimately to $c_{\rm eff}(\kappa_+,\kappa_-)=(1-2\kappa_{\rm max})\ln 2$ with $\kappa_{\rm max}=\max(\kappa_+,\kappa_-)$.
If the defect is irrelevant on both sides, i.e. $\kappa_+,\kappa_->1/2$, it acts as a local defect for the entanglement entropy, and the effective central charge is expected to be an increasing function of both $\kappa_+$ and $\kappa_-$.   
Finally, if the defect is relevant in exactly one subsystem, i.e. either 
$\kappa_+<1/2$ and $\kappa_->1/2$, or $\kappa_+>1/2$ and $\kappa_-<1/2$, the arguments of Ref. \cite{jkri} lead to a double-logarithmic scaling of the entanglement entropy $S_L\sim \ln\ln L$. 

Lastly, we mention that it would be interesting to study the behaviour of entanglement measures other than the entanglement entropy in the present model, such as the logarithmic negativity, which has recently been studied in random singlet phases \cite{ruggiero,shapourian}.

\ack
The author thanks F. Igl\'oi for useful comments.
This work was supported by the Hungarian Scientific Research Fund under Grant
No. K109577.


\end{document}